\documentclass[aip,jcp,reprint,floatfix,a4paper]{revtex4-1}

\usepackage{graphicx}
\usepackage[sort&compress]{natbib}
\usepackage[sumlimits]{amsmath}
\usepackage{amsfonts}
\usepackage{amssymb}
\usepackage{dcolumn}
\usepackage{subdepth}
\usepackage{color}
\usepackage{enumitem}
\usepackage{hyperref}
\usepackage{blindtext}
\usepackage{subdepth}

\begin{document}
\title{Herman-Kluk propagator is free from zero-point energy leakage}

\author{Max Buchholz}
% \affiliation{Institut f\"ur Theoretische Physik, Technische Universit\"at Dresden, 01062 Dresden, Germany}
\affiliation{Dipartimento di Chimica, Universit\`a degli Studi di Milano, via Golgi 19, 20133 Milano, Italy}
\author{Erika Fallacara}
\affiliation{Dipartimento di Chimica, Universit\`a degli Studi di Milano, via Golgi 19, 20133 Milano, Italy}
\author{Fabian Gottwald}
\affiliation{Institute of Physics, Rostock University, Albert-Einstein-Str. 23-24, 18059 Rostock, Germany}
\author{Michele Ceotto}
\affiliation{Dipartimento di Chimica, Universit\`a degli Studi di Milano, via Golgi 19, 20133 Milano, Italy}
\author{Frank Grossmann}
\email{frank@physik.tu-dresden.de}
\affiliation{Institut f\"ur Theoretische Physik, Technische Universit\"at Dresden, 01062 Dresden, Germany}
\author{Sergei D. Ivanov}
\email{sergei.ivanov@uni-rostock.de}
\affiliation{Institute of Physics, Rostock University, Albert-Einstein-Str. 23-24, 18059 Rostock, Germany}

%Macro employed in the text. PLEASE USE!
\newcommand{\Eq}[1]{Eq.\,(\ref{#1})}
\newcommand{\Eqs}[1]{Eqs.\,({#1})}
\newcommand{\Sec}[1]{Sec.\,\ref{#1}}
\newcommand{\e}{\mathrm{e}}
\newcommand{\im}{\mathrm{i}}
\newcommand{\diff}{\mathrm{d}}
\newcommand{\cc}[1]{\textcolor{red}{#1}} % critical comments
\newcommand{\Fig}[1]{Fig.\,\ref{#1}}
\newcommand{\cm}{cm$^{-1}$}

\newcommand{\Ref}[1]{Ref.~#1}
\newcommand{\aumath}{\text{a.u.}}

\begin{abstract}
%SI: Abstract reworked. To start with setting the stage.
Semiclassical techniques constitute a promising route to approximate quantum dynamics based on classical trajectories starting from a quantum-mechanically correct distribution.
One of their main drawbacks is the so-called zero-point energy (ZPE) leakage, that is artificial redistribution of energy from the modes with high frequency and thus high ZPE to that with low frequency and ZPE due to classical equipartition.
%SI: We don't really study the appearance of the leakage.
%We investigate the appearance as well as the leakage of zero point energy in systems of a few coupled oscillators. In classical calculations, leakage of zero point energy has been observed if it is putinto the system initially. A similar set of trajectories, however, leads to the correct behaviour, i.e., no leakage, if it is used in 
Here, we show that 
an elaborate semiclassical formalism based on the Herman-Kluk propagator is free from the ZPE leakage despite utilizing purely classical propagation.
This finding opens the road to correct dynamical simulations of systems with a multitude of degrees of freedom that cannot be treated fully quantum-mechanically due to the
exponential increase of the numerical effort.
\end{abstract}

\date{\today}
\maketitle

\section{Introduction\label{sec:introduction}}
%----------------------------------------------------------

Understanding dynamical processes happening in complex many-body molecular systems is one of the main tasks of modern theoretical chemistry.
To-date simulation approaches allow one to bridge the gap between theoretical models and experiments, shedding light onto the processes in question on the microscopic level.
In this context, classical (\textit{ab initio}) molecular dynamics (MD) methods enjoyed success during the last few decades, owing to their utmost robustness and simplicity.~\cite{Marx-Book,Tuckerman-Book}
Nonetheless, the presence of quantum coherence, light atoms, shallow potential energy surfaces (PESs), low temperatures and/or isotope substitutions may lead to a qualitatively wrong behavior if the nuclei are treated classically, 
as was shown on numerous examples starting from small molecules in gas phase to biomolecules.~\cite{Gao-AR-2002,Olsson-JACS-2004,Habershon-JCP-2009,Ivanov-NatChem-2010,Witt-PRL-2013}
In particular, the importance of 
%SI: Now introduced in the abstract
the ZPE
%zero-point energy (ZPE)
in the context of elementary atom-transfer reactions was realized from the early days of the simulation era and quasiclassical trajectory (QCT) methods, utilizing purely classical simulations started from the correct quantum distribution, have emerged, see e.g.\ Refs.~\onlinecite{Truhlar-JPC-1979,Schatz-JCP-1983} and references therein.
These QCT methods are suffering from the so-called ZPE leakage, that is the energy flow from high-frequency modes to low-frequency ones due to equipartition, as was also 
realized from early on.~\cite{Bowman-JCP-1989,Miller-JCP-1989}
Various attempts to circumvent this problem included using reduced models~\cite{Haze-JCP-1988} and constraints that prevent the vibrational energy in a mode from falling below its zero-point value.~\cite{Bowman-JCP-1989,Miller-JCP-1989,Peslherbe-JCP-1994,Lim-JCP-1995}
Further, methods sacrificing those trajectories that do not satisfy the ZPE criterion~\cite{Nyman-JCP-1990,Varandas-JCP-1992} emerged as well as that based on an $N$-mode representation of the coupling part in the potential with smooth elimination of the terms as the energy of any mode falls below a specified value corresponding to ZPE~\cite{Bowman-JPCA-2006} to mention but few.
Some of these developments were debated, questioning the whole concept of excluding the regions of phase space where ZPE constraint does not hold,~\cite{Schlier-JCP-1995,McCormack-JCP-1995} and systematically investigated~\cite{Sewell-CPL-1992,Guo-JCP-1996} ``\ldots with the conclusion that nearly all of the approaches that have been proposed are unfounded and aphysically affect the dynamics''.~\cite{Guo-JCP-1996}
In the last decade these ideas were revived in the context of efficient thermostatting, and the colored-noise thermostat that yields quantum-mechanically correct momentum and position distributions was established.~\cite{Ceriotti-PRL-2009}
%
%SI: Actually, the difference between the two is not that large, is it.
%FG: I think the QTB method is just a normal Langevin thermostat isn't it? So it has no memory kernel but just some quantum FDT
%SI: OK, we should look at the equations of QTB again to figure that out with certainty.
%FGII: issue might be solved by Finocci......
Few months later, a similar idea was independently developed and termed ``quantum thermal bath'' (QTB).~\cite{Dammak-PRL-2009}
These methods were successful for systems where the degree of anharmonicity was not very high and, although they are prone to the ZPE leakage, it was shown that choosing the coupling strength of a thermostat strong enough remedies the problem at least for static properties.~\cite{Brieuc-JCTC-2016-ZPE}
Nonetheless, having such a strong coupling leads to, e.g.,
%SI: added
artificially
 broadened spectral lineshapes as was shown therein as well.
Thus, a necessity for a more systematic and fundamental approach to the problem became apparent.

From a historic perspective, there exist two classes of methods achieving this goal.
The first one unites the imaginary-time path integral (PI) approaches, based on the Feynman PIs~\cite{Feynman-Book-1965} and the associated ``classical isomorphism''.~\cite{Chandler-JCP-1981,Tuckerman-Book}
The latter connects the partition function of a quantum particle to a configurational integral of some more complicated but purely classical object.
It has the form of a beaded necklace, with adjacent beads connected with harmonic springs, and is often referred to as the ring polymer.
Simulating it via MD or Monte Carlo methods allows for genuine quantum effects such as the ZPE, yielding numerically exact results for static (thermodynamic) properties.
Unfortunately, any real-time information may be obtained in an approximate fashion only.
Here, the ring polymer molecular dynamics~\cite{Craig-JCP-2004} (RPMD) method became increasingly popular, see e.g.\ Refs.~\onlinecite{Habershon_ARPC_2013,Ceriotti-CR-2016} for review.
It delivers classical dynamics that naturally preserves the quantum Boltzmann density, though no information about the phases is available, leaving quantum coherence effects outside reach.
Furthermore, when it comes to vibrational (infrared) spectroscopy, RPMD suffers from artificial resonances
of the aforementioned springs with the system modes.~\cite{Witt-JCP-2009,Habershon-JCP-2008}
Although the problem was mitigated by attaching a tailored Langevin thermostat that removes the resonances due to its stochastic nature,~\cite{Rossi-JCP-2014} it is still not the ultimate solution, 
as it may affect the dynamics of the system in an undesired way.

The second class comprises the so-called semiclassical methods, that emerged from the propagator suggested by van Vleck back in 1928, which was exclusively based on classical trajectories.~\cite{VanVleck-PNAS-1928, Littlejohn1992}
A necessary ingredient to utilize MD simulation techniques is the initial-value representation (IVR) that recasts the problem into a propagation of an initial (quantum) distribution in phase space, see Refs.~\onlinecite{Miller-ACP-1974,Miller-JPCA-2001,Thoss-AR-2004,Kay-AR-2005} for
reviews.
Expanding the Heisenberg evolution operator in the coherent states' basis led to the Herman-Kluk (HK) propagator,~\cite{Herman-Kluk-CP-1984, Herman-Kluk-JCP-1986,Kay-JCP-1994b,Grossmann-Book-2013} which can be 
viewed as a frozen-Gaussian IVR approximation.~\cite{Heller-JCP-1981,Wehrle-Vanicek-JCP-2014}
Alternatively, semiclassical propagators can be formulated based on the Wigner formulation of quantum mechanics,~\cite{DeAlmeida-PR-1998} see Refs~\onlinecite{Dittrich-PRL-2006,Dittrich-JCP-2010,DeAlmeida2013} for selected representatives.
Very recently Koda suggested a universal recipe to formulate semiclassical Wigner propagators based on the existing Hilbert-space ones,~\cite{Koda-JCP-2015} via the analogy of the Moyal equation for the Wigner function and the Schr\"odinger equation.~\cite{Bondar-PRA-2013}
Thereby, the Wigner version of the HK propagator was suggested for the first time and the Wigner counterpart of the van Vleck propagator~\cite{Dittrich-PRL-2006,Dittrich-JCP-2010} was re-derived and re-formulated in terms of an IVR.
On this basis, some of us have presented a unified viewpoint on the van Vleck and HK propagators in Hilbert space and in Wigner representation.~\cite{Gottwald-CP-2018}
According to it, the Wigner HK propagator is conceptually the most general one although it has no performance benefits over its well-established Hilbert-space counterpart.
Most of other semiclassical propagators are its limiting (and non-optimal) cases and, thus, practical applications are mostly based on the HK propagator in Hilbert space.
%, i.e.\ using it as a propagator for a wavefunction or incorporating it into the superoperator for a density matrix propagation.
%
Since it usually suffers from the infamous sign problem which is caused by rapid oscillations in phase factors, several approximations have been developed based on (modified) Filinov 
filtering,~\cite{Makri-CPL-1987, Walton-MolPhys-1996,Herman-CPL-1997} time-averaging 
methods~\cite{Elran-JCP-1999a, Elran-JCP-1999b,Kaledin-JCP-2003, Ceotto-JCP-2009, Gabas_Ceotto_Glycine_2017, ceotto_conte_DCSCIVR_2017, DiLiberto_Ceotto_Jacobiano_2018} or forward-backward schemes~\cite{Sun-JCP-1999,Makri-CPL-1998} allowing 
one to deal with systems of up to hundreds of degrees of freedom (DOFs) in various 
contexts.~\cite{Kuehn-JPCA-1999,Batista-JCP-1999, Skinner-JCP-1999,Guallar-JCP-2000,Wang-JCP-2000,Ovchinnikov-JCP-2001,Nakayama-JCP-2003,Nakayama-CP-2004,Tao-JCP-2009,Herman-JPCB-2014,Alemi-JCP-2015}
Other approaches employ improved sampling techniques~\cite{Tao-JCP-2009,Tao-JCP-2014} or hybrid schemes treating certain unimportant DOFs less accurately~\cite{Grossmann-JCP-2006,Buchholz-JCP-2016, Ceotto_Buchholz_MixedSC_2017, Ceotto_Buchholz_SAM_2018} or even implicitly as a heat bath.~\cite{Koch-PRL-2008,Koch-CP-2010}
Even further simplification led to the so-called Wigner model, also referred to as the linearized
semiclassical initial-value representation (LSC-IVR).~\cite{Heller-JCP-1976, Miller-JCP-1998, Liu_Linearized_2015}
Unfortunately, the latter one was shown to suffer from the ZPE leakage problem even in condensed phase~\cite{Habershon-JCP-2009} as it is exclusively based on the classical propagation, 
and the question arises if also 
%SI: 
truly
semiclassical methods suffer from this problem.

%SI: paragraph slightly polished 
As a central result, we will show that the fully semiclassical HK propagator, although being exclusively based on classical trajectories, is free from the ZPE leakage.
To this end,
%SI:
% in the following,
we first recapitulate semiclassical and classical approaches for the simulation of expectation values in \Sec{sec:theory}.
Then the 2D and 3D model oscillator systems
%SI: 
tailored to exhibit strong ZPE leakage
% exhibiting a 2:1 resonance,
are introduced in \Sec{sec:model}.
Semiclasscial and classical results for the ZPE conservation, respectively leakage, are presented and discussed in \Sec{sec:results}.

%----------------------------------------------------------
\section{Theory\label{sec:theory}}
%----------------------------------------------------------

We first give a brief overview on the semiclassical and linearized semiclassical 
%SI: I think a more conventional term is 
simulation techniques and present working expressions particularly for energy expectation values
%calculation of expectation values of the energy that
that are needed to elaborate on the ZPE conservation
% we need in order to see if zero point energy is conserved
in any of the two approaches.

%SI: I still didn't get why it is \rho_\alpha
For an $F$-dimensional system
%SI: moved from below, CHECK ME. And somehow \hat{\rho}_\alpha(0) is still not introduced and its dependence on \alpha is not clear
and for an initial state of Gaussian form centered around $({\bf p}_\alpha,{\bf q}_\alpha)$,
the reduced density matrix based on the HK propagator~\cite{Herman-Kluk-CP-1984} reads 
\begin{align}
\rho_{\alpha,i}\left(x_i,x_i',t\right)=
&\int\frac{\diff^F p_0\diff^F p_0'\diff^F q_0\diff^F q_0'}{(2\pi\hbar)^{2F}} 
\nonumber
\\
&\times C_{t}(C_{t}')^{*}\text{e}^{\text{i}(S_{t}-S_{t}')/\hbar}
\nonumber
\\
&\times\left\langle g\left(\mathbf{p}_{\text{e},t}',\mathbf{q}_{\text{e},t}'\right)\middle|g\left(\mathbf{p}_{\text{e},t},\mathbf{q}_{\text{e},t}\right)\right\rangle 
\nonumber
\\
&\times\left\langle g\left(p_{i,t}',q_{i,t}'\right)\middle|x_i'\right\rangle \left\langle x_i\middle|g\left(p_{i,t},q_{i,t}\right)\right\rangle 
\nonumber
\\
&\times\left\langle g\left(\mathbf{p}_0,\mathbf{q}_0\right)\middle|\hat{\rho}_\alpha(0)\middle|g\left(\mathbf{p}'_0,\mathbf{q}'_0\right)\right\rangle 
\label{eq:rho_HK} 
\enspace ,
\end{align}
whose main ingredients are Gaussian wavepackets
\begin{align}
\langle \mathbf{x}|g(\mathbf{p},\mathbf{q})\rangle=&\left(\frac{\det{\boldsymbol{\gamma}}}{\pi^F}\right)^{1/4}
\exp\left\{-\frac{1}{2}(\mathbf{x}-\mathbf{q})\cdot{\boldsymbol{\gamma}}(\mathbf{x}-\mathbf{q})\right .
\nonumber
\\
&+\left .\frac{{\rm i}}{\hbar} \mathbf{p}\cdot(\mathbf{x}-\mathbf{q})\right\}
\end{align}
with a fixed width-parameter diagonal matrix
$\boldsymbol{\gamma}$ and $S_t$ being the classical action along the trajectory. 
The expression in \Eq{eq:rho_HK} is obtained by choosing a single Cartesian DOF $x_i$ and tracing the full density matrix 
$\rho_{\alpha}\left(\mathbf{x},\mathbf{x}',t\right)$
over the remaining $(F-1)$-environmental DOFs,
%SI: The sentences are united
%The $(F-1)$-dimensional environmental component of the trajectories is 
collectively denoted by the subscript ``$\text{e}$''.
This reduced density matrix is based on the classical trajectories, which start at $(\mathbf p_{0},\mathbf q_{0})$ at time $t=0$ and reach the phase-space point $(\mathbf p_{t},\mathbf q_{t})=[(p_{i,t},\mathbf  p_{\text{e},t}),(q_{i,t},\mathbf q_{\text{e},t})]$ at time $t$.
The preexponential weight factor of such a trajectory in phase space is given by
\begin{equation}
\label{eq:prefac}
C_t=\sqrt{\frac{1}{2^F}\det\bigg({\mathbf m}_{\mathbf{pp}}+{\mathbf m}_{\mathbf{qq}}
-{\rm i}\hbar \boldsymbol{\gamma} {\mathbf m}_{\mathbf{qp}}-\frac{1}{{\rm i}\hbar}{\mathbf m}_{\mathbf{pq}}\boldsymbol{\gamma}^{-1}\bigg)}
\enspace ,
\end{equation}
which is composed of the four block matrices of the monodromy matrix, that are defined as
%SI:
${\mathbf m}_{\mathbf{ab}}\equiv \partial {\mathbf a}_t/\partial {\mathbf b}_0, \forall a,b \in \{q,p\}$.
%${\mathbf m}_{\mathbf{pp}}\equiv \partial {\mathbf p}_t/\partial {\mathbf p}_0$ etc.
%
In a numerical implementation the integration is replaced by a sum and
the semiclassical contribution of a single trajectory 
is then weighted by $C_t {\rm e}^{{\rm i} S_t/\hbar}$.
Convergence is achieved with a finite number of trajectories through the overlap between the initial state
and the Gaussian.
For a review of this semiclassical IVR (SC-IVR) methodology and related approaches, see \Ref{\onlinecite{Grossmann-CAMP-1999}}.
%
%SI:
In order to obtain
%For the simulation of
expectation values, one has to deal with double phase-space integrals,
which are treated using the combined sampling strategy as laid out in \Ref{\onlinecite{Lasser-Sattlegger-NM-2017}}.

As it was discussed in the Introduction, the method that was shown to suffer from the ZPE leakage is LSC-IVR also referred to as the Wigner model.
It is based on generalized correlation functions comprising two time-evolution operators such that a double Herman-Kluk
expression emerges in a full semiclassical description. 
A linear expansion of the action difference leads to a purely classical expression in terms of a single ($2F$-dimensional) 
phase-space integral with no quantum interference effects. 
Heller had written down the result for the correlation function intuitively,~\cite{Heller-JCP-1976}
whereas semiclassical derivations have been given by Miller and coworkers~\cite{Miller-JCP-1998} as well as
by Herman and Coker.~\cite{Herman-Coker-JCP-1999}
In this framework,
the diagonal elements of the reduced density matrix in position representation are given by
\begin{align}
\rho_{\alpha,i}({x_i},t)=& \int\frac{\diff^F p_0\diff^F q_0}{(\pi\hbar)^F}\delta [q_{i,t}-x_i] 
\nonumber 
\\ 
&\times {\rm e}^{-({\bf p}_0-{\bf p}_\alpha)\cdot\boldsymbol{\gamma}^{-1}({\bf p}_0-{\bf p}_\alpha)/\hbar^2-({\bf q}_0
-{\bf q}_\alpha)\cdot\boldsymbol{\gamma} ({\bf q}_0-{\bf q}_\alpha)}
\enspace ,
\label{eq:lsc_pos}
\end{align}
which contains the system part of the classical trajectories $q_{i,t}$ as the only dynamical input.
The momentum space analogue reads
\begin{align}
\rho_{\alpha,i}(p_{x,i},t)=& \int\frac{\diff^F p_0\diff^F q_0}{(\pi\hbar)^F}\delta [p_{i,t}-p_{x,i}] 
\nonumber 
\\ 
&\times {\rm e}^{-({\bf p}_0-{\bf p}_\alpha)\cdot\boldsymbol{\gamma}^{-1}({\bf p}_0-{\bf p}_\alpha)/\hbar^2-({\bf q}_0
-{\bf q}_\alpha)\cdot\boldsymbol{\gamma} ({\bf q}_0-{\bf q}_\alpha)}
\enspace,
\label{eq:lsc_mom}
\end{align}
where $p_{x,i}$ designates the momentum conjugate to coordinate $x_i$.
The LSC-IVR method is taking into account the full quantum nature of the initial state
but apart from that is purely classical and thereby cannot describe any interference effects.

The main quantities of interest are the individual energies of each site, defined as
\begin{align}
 \label{eq:energies}
 E_i(t) = \frac{1}{2} \left\langle \hat{p}_{x,i}^2 \right\rangle(t) + \frac{\omega_i^2}{2} \left\langle \hat{x}_i^2 \right\rangle(t)
\end{align}
for the model
%SI:
system
of coupled harmonic oscillators, see \Sec{sec:model}.
%
%SI:
Within the HK and LSC-IVR approaches, one
%With HK and LSC-IVR, we 
can find analytical expressions for the expectation values, 
%SI:
in order to
%such that we can
circumvent calculating the reduced density matrix.
In the HK case, 
%SI: I think it reads better, when not personal. Changed throughout
the second moment of the $i$th coordinate can be found with the help of \Eq{eq:rho_HK} as 
%we employ \Eq{eq:rho_HK} and find the second moment of position of the $i$th coordinate as 
%
\begin{align}
 \label{eq:q2_HK}
 \left\langle \hat{x}_i^2 \right\rangle(t)=&\ \text{Tr}\left(\hat{x}^2_i\hat{\rho}_{\alpha,i}(t)\right)
 \nonumber
 \\
 =&\int \diff x_i x_i^2 \rho_{\alpha,i}\left(x_i,x_i,t\right)
 \nonumber
 \\
 =&\int\frac{\diff^F p_0\diff^F p_0'\diff^F q_0\diff^F q_0'}{(2\pi\hbar)^{2F}} C_{t}\left(C_{t}'\right)^{*}\text{e}^{\text{i}(S_{t}-S_{t}')/\hbar}
\nonumber
\\
&\times\left\langle g\left(\mathbf{p}_{t}',\mathbf{q}_{t}'\right)\middle|g\left(\mathbf{p}_{t},\mathbf{q}_{t}\right)\right\rangle 
\nonumber
\\
&\times\left\langle g\left(\mathbf{p}_0,\mathbf{q}_0\right)\middle|\hat{\rho}_\alpha(0)\middle|g\left(\mathbf{p}'_0,\mathbf{q}'_0\right)\right\rangle 
\nonumber
\\
&\times \frac{1}{2\gamma_i}\left( 1+\frac{D_{i,+}^2}{2\gamma_{i}} \right)
\enspace ,
\end{align}
where the abbreviation $D_{i,+}$ stands for the coefficient of the first order term in the exponent of the Gaussian integrand
\begin{align}
 \label{eq:expectations_HK_abbrev}
 D_{i}&\equiv  \left(\gamma_i q_{i,t}+\frac{\rm i}{\hbar}p_{i,t}\right)
 \\
 D_{i,\pm}& \equiv D_{i}\pm \left(D_{i}'\right)^*.
\end{align}
The calculation of the momentum expectation value works analogously
\begin{align}
 \label{eq:p2_HK}
 \left\langle \hat{p}_{x,i}^2 \right\rangle(t)=&\ \text{Tr}\left(\hat{p}^2_{x,i}\hat{\rho}_{\alpha,i}(t)\right)
 \nonumber
 \\
 =&\int \diff x_i \left. \frac{\partial^2}{\partial x_i'^2} \rho_{\alpha,i}\left(x_i,x_i',t\right) \right|_{x_i=x_i'}
 \nonumber
 \\
 =&\int\frac{\diff^F p_0\diff^F p_0'\diff^F q_0\diff^F q_0'}{(2\pi\hbar)^{2F}} C_{t}\left(C_{t}'\right)^{*}\text{e}^{\text{i}(S_{t}-S_{t}')/\hbar}
\nonumber
\\
&\times\left\langle g\left(\mathbf{p}_{t}',\mathbf{q}_{t}'\right)\middle|g\left(\mathbf{p}_{t},\mathbf{q}_{t}\right)\right\rangle 
\nonumber
\\
&\times\left\langle g\left(\mathbf{p}_0,\mathbf{q}_0\right)\middle|\hat{\rho}_\alpha(0)\middle|g\left(\mathbf{p}'_0,\mathbf{q}'_0\right)\right\rangle 
\nonumber
\\
&\times \frac{\hbar^2\gamma_i}{2}\left( 1-\frac{D_{i,-}^2}{2\gamma_i} \right)
\enspace .
\end{align}

In the classical case, combining the two LSC-IVR expressions in \Eq{eq:lsc_pos} and \Eq{eq:lsc_mom} yields the simple formula
\begin{align}
 \label{eq:lsc_energy}
 E_i(t) = 
 & \int\frac{\diff^F p_0\diff^F q_0}{(\pi\hbar)^F}\left(\frac{1}{2}p_{i,t}^2+\frac{\omega_i^2}{2}q_{i,t}^2\right)
\nonumber 
\\ 
&\times {\rm e}^{-({\bf p}_0-{\bf p}_\alpha)\cdot\boldsymbol{\gamma}^{-1}({\bf p}_0-{\bf p}_\alpha)/\hbar^2-({\bf q}_0
-{\bf q}_\alpha)\cdot\boldsymbol{\gamma} ({\bf q}_0-{\bf q}_\alpha)}
\end{align}
for the energy of the $i$th site.
Results obtained via both approaches are presented in \Sec{sec:results}.

%----------------------------------------------------------
\section{Model and technical details \label{sec:model}}
%----------------------------------------------------------

%SI: I changed the technical details to simple past, as we are writing about what was done
In order to illustrate the difference between the LSC-IVR and HK
%SI: 
methods
% results
with respect to ZPE leakage, a simple model was employed that meets two requirements: i) there must be a ZPE leakage as a result of the classical propagation; ii) the model should be feasible for the HK method to yield converged results.
%
%SI: sentence reworked
To this end, a system consisting of two cubically coupled harmonic oscillators utilized in \Ref{\onlinecite{Brieuc-JCTC-2016-ZPE}} was generalized to more than two DOFs as
\begin{align}
 \label{eq:Hamiltonian}
 H = \sum\limits_{i=1}^F \frac{p_i^2}{2} +
     \sum\limits_{i=1}^F \frac{\omega_i^2 q_i^2}{2} + 
     \sum\limits_{i<j} C_{ij} \left(q_i-q_j\right)^3
 \enspace ,
\end{align}
%
%SI: I think the fact that oscillators have mass one has nothing to do with atomic units!
where all oscillators had unity mass and atomic units were used, that is 
%In atomic units, all oscillators have mass one and
$\hbar=1$.
The frequencies of all but one oscillator were taken to be $\omega_i=0.01\,\aumath$, $i=1,\dots,F-1$, while the frequency for the remaining oscillator was set to
%SI:
$\omega_F=0.005\,\aumath$
%$\omega_j=0.005\,\aumath$, $j=F$.
%
The coupling strength was chosen as $C_{12}=10^{-8}\,\aumath$ for the case of two harmonic oscillators. 
For the three-dimensional simulations, two different coupling strengths between low-frequency and high-frequency oscillators were employed, $C_{13}=10^{-8}\,\aumath$ and $C_{23}=2\times 10^{-8}\,\aumath$, while the coupling between the two high-frequency oscillators was $C_{12}=10^{-8}\,\aumath$
As discussed in \Ref{\onlinecite{Brieuc-JCTC-2016-ZPE}}, cubic coupling terms typically cause overtone frequencies and, due to the 2:1 frequency ratio, all high-frequency oscillators are in resonance with the
%SI:
first
overtone of the low-frequency oscillator.
This resonant energy flow in the classical propagation
%SI:
had been
made responsible for the observed ZPE leakage in \Ref{\onlinecite{Brieuc-JCTC-2016-ZPE}} and, thus, we view this setup as an optimal test case
%SI:
for the present purpose as well.
%for checking whether the HK propagator suffers from the ZPE leakage as well.

The initial state was
%SI:
chosen as 
a product of Gaussian wavepackets with widths $\gamma_i=\omega_i$, centered at $(\mathbf{p}_{\alpha},\mathbf{q}_{\alpha})=(\mathbf{0},\mathbf{0})$, which means that each site initially had exactly the ZPE.
Each initial state was propagated with
%SI: TODO: to specify which one
a symplectic integrator for 5000 steps of length $10\,\aumath$, resulting in a total propagation time that approximately corresponds to 40 periods of the low-frequency oscillator.
%
%SI: sentence polished
Exact quantum calculations performed via the split-operator FFT, implemented in the WavePacket software,\cite{Schmidt2017-WavePacket1,Schmidt2018-WavePacket2} were used a reference.
While a single phase-space integration in the LSC-IVR simulation according to \Eq{eq:lsc_energy} required only
%SI:
few dozen
thousand trajectories, the HK simulation was much more demanding. 
This is, on one hand, due to the double rather than single 
%SI:?
phase-space
%phase
integration in \Eq{eq:q2_HK} and \Eq{eq:p2_HK}, and, on the other hand, due to the more complicated form of the phase-space integrand itself. 
Thus, 10 million trajectories were needed to achieve convergence for the propagation time of 50000\,a.u. for the 2D system. 
For the 3D case, 40 million trajectories sufficed for good convergence until $\approx 25000\,\aumath$
%
%SI: polished 
Since going beyond three harmonic oscillators required even larger number of trajectories in order to obtain the long-time convergence, we didn't not include these results here. 
%
% SPECTRAL CONSIDERATIONS REMOVED
%
% While convergence is a limiting factor for the maximum useful propagation time, in particular for the double phase space integration that is needed for the energy expectation values, 
% it is much more easily achieved for spectral considerations if the time-averaging formalism is employed. To this end, we propagate $10^5$ trajectories for a total 
% propagation time of $2.06\times 10^{6}\ \text{a.u.}$ to arrive at a frequency resolution of $3\times10^{-6}\ \text{a.u.}$  

%----------------------------------------------------------
\section{Results and Discussion\label{sec:results}}
%----------------------------------------------------------

\begin{figure}
 \includegraphics[width=\columnwidth]{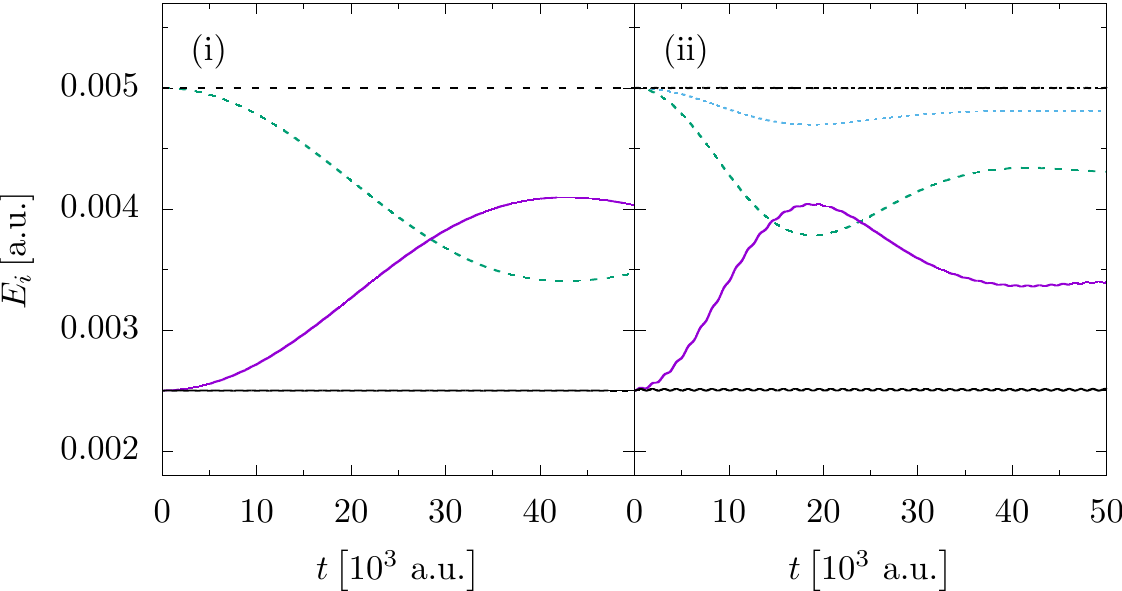}
 \caption{Energies of individual sites. From left to right: 2D (i) and 3D results (ii). Solid lines: energy of the low-frequency oscillator (violet: LSC-IVR, black: FFT), dashed lines: energies of the high-frequency oscillators (turqoise and light blue: LSC-IVR, black: FFT).
          \label{fig:united}}
\end{figure}

First, let us consider the site energies computed with the aforementioned methods.
As can be seen in \Fig{fig:united}, site energies computed from the LSC-IVR method show a strong ZPE leakage for both 2D and 3D systems.
% 
%SI: This is so by construction, right? It is not an achievement of the method.
While all oscillators start at the exact ZPE,
%SI: CHECK ME
as it is ensured by the choice of the initial state, 
% i.e., in their respective ground states (the $|0,0\rangle$-state),
the subsequent evolution of these energies using LSC-IVR deviates considerably from the exact results, which preserve the ZPE for all times.
In particular, energy is dissipated from the high-frequency mode(s) into the low-frequency mode, regardless of the number of DOFs. We note in passing that a 
related study has been performed for the breather initial condition $|1,0\rangle$, i.e., with one oscillator in its first excited state and the second one in its ground state.\cite{Zagoya-2014}

\begin{figure}
 \includegraphics[width=\columnwidth]{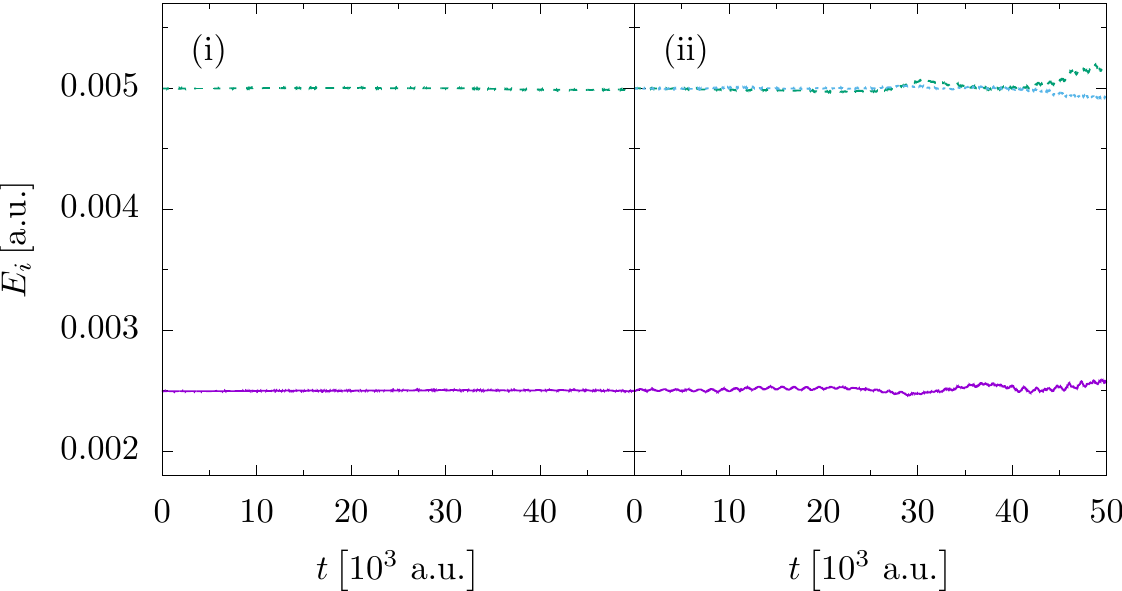}
 \caption{Energies of individual sites, calculated with HK. From left to right: 2D (i), 3D (ii).
          Solid violet line: energy of the low-frequency oscillator, all dashed lines: energies of the high-frequency oscillators.
          \label{fig:HK_2D_to_3D}}
\end{figure}

Having affirmed that a simple averaging over classical trajectories starting from a quantum initial state is not sufficient to prevent the ZPE leakage, we come to the question at the heart of this investigation: is there still a ZPE leakage in a truly semiclassical method that allows for the interference of different trajectories?
The answer can be found in the results of the HK simulations, see \Fig{fig:HK_2D_to_3D}. 
%SI:
The energies of 
both high- and low-frequency oscillators 
%SI:
remain almost constant during the entire propagation time.
%stay close to their respective initial energies.
This is 
%SI:
especially true
%particularly obvious
for the two-dimensional system, where the quantum result is reproduced almost exactly.
%SI: Is it really convergence and not the deficiency of HK as such? Can you display error bars please?
%MB and FGII: if time is left after calultion of velocity velocity....
The deviations seen in the three-dimensional case, in particular towards the end of the propagation, may be attributed to insufficient convergence of the HK results. 

All in all, the simple model investigation presented above reveals that the HK propagator, despite relying on classical trajectories, is free from the ZPE leakage.
This is a consequence of its elaborate structure which interconnects the trajectories in a highly non-trivial way.
Although the considered systems are very simple, the demonstrated absence of the ZPE leakage cannot be a consequence of this simplicity.
In contrast, this result should hold for arbitrary systems and thus the goal is to find a proper approximation to the method that preserves the advantages of the HK propagator and circumvents the numerical weaknesses it is suffering from.

\section{Acknowledgements}

Michele Ceotto and Max Buchholz acknowledge financial support from the European Research Council (ERC) under the European Union’s Horizon 2020 research and innovation programme (Grant Agreement No.~[647107] -- SEMICOMPLEX -- ERC-2014-CoG). M.C.\ acknowledges also the CINECA for the availability of high performance computing resources.

\bibliography{./united}
\end{document}